\documentclass[final,3p,times,twocolumn]{elsarticle}

\usepackage{graphicx}
\usepackage{graphicx}
\usepackage{multicol,lipsum}
\usepackage{amsmath}
\usepackage{dcolumn} 
\usepackage{enumitem}
\usepackage{graphicx}
\usepackage{dcolumn}
\usepackage{array,multirow}
\usepackage{bm}
\usepackage{amsmath}
\usepackage{epstopdf}
\usepackage{amsfonts}
\usepackage{amssymb}
\usepackage{mathrsfs}
\usepackage{epsfig}
\usepackage{tabularx}
\usepackage{color}
\usepackage[dvipsnames]{xcolor}

\newcommand{\ed}{\mathrm{d}}

\newcommand{\rev}{\mathrm{rev}}

\journal{Physics Letters B}
\usepackage{hyperref}
\hypersetup{
    colorlinks = true}

\begin{document}

\begin{frontmatter}

\title{Adiabatic evolution of Hayward black hole}

\author[label1]{Mohsen Fathi}
\ead{mohsen.fathi@postgrado.uv.cl}
\author[label2]{Mart\'in Molina}
\ead{martin.molina.r@mail.pucv.cl}
\author[label1]{J.R. Villanueva}
\ead{jose.villanueva@uv.cl}

\address[label1]{Instituto de F\'isica y Astronom\'ia, Universidad de Valpara\'iso, Avenida Gran Breta\~na 1111, Valpara\'iso, Chile\fnref{label1}}

\address[label2]{Instituto de F\'isica, Pontificia Universidad Cat\'olica de Valpara\'iso, Avenida Brasil 2950, Valpara\'iso, Chile\fnref{label2}}

\begin{abstract}

In this letter we use the Carath\'{e}odory's approach to thermodynamics, to construct the thermodynamic manifold of the Hayward black hole. The Pfaffian form representing the infinitesimal heat exchange reversibly is considered to be $\delta Q_{\mathrm{rev}}\equiv {\rm d}r_s-\mathcal{F}_H {\rm d}l$, previously obtained by Molina \& Villanueva \cite{fmv20}, where $r_s$ is the Schwarzschild radius, $l$ is the Hayward's parameter responsible for the possible regularization of the Schwarzschild black hole, and $\mathcal{F}_H$ is the intensive variable called the Hayward's force.
By solving the associated Cauchy problem, the adiabatic paths are confined to the non-extremal manifold, and therefore, the status of the second and third laws are preserved. 
Consequently, the extremal sub-manifold corresponds to the {adiabatically disconnected} boundary of the manifold.
In addition, the merger of two extremal Hayward black holes is analyzed. 
\end{abstract}


\end{frontmatter}


\section{Introduction}

{
Ever since their advent, the reconciliation of the laws of thermodynamics with black hole mechanics \cite{Bardeen:1973gs}, the entropy assigned by Bekenstein to the black holes \cite{Bekenstein:1972,Bekenstein:1973,Bekenstein:1974,Bekenstein:1975} and the possibility of black hole evaporation through the Hawking radiation \cite{Hawking:1974sw}, have been of great interest among physicists. And although it has not been possible to detect such phenomena from direct observations, nevertheless, strong effort have been being made to mimic similar processes in black hole analogs, such as the experimental Unruh radiation \cite{Unruh:1981} in stimulated systems, both theoretically and experimentally \cite{Novello:2002,Schutzhold:2005,Carusotto:2008,Belgiorno:2010,Weinfurtner:2011,Castelvecchi:2016,Steinhauer:2016,Lima:2019,Kolobov:2021}. On the other hand, while the famous Bekenstein-Hawking (B-H) entropy formula has been applied widely for the regular black holes, nevertheless, its direct application to the extremal black holes (EBHs) is not that simple.} 
In fact, the special conjecture of zero entropy for EBHs \cite{Teitelboim:1994,Carroll:2009}, leads to overlooking the direct relationship between the entropy and the event horizon's area, as demanded by the B-H formula.

Along with the strong interest of the community in the study of the thermodynamics of regular black holes, in this paper, we focus on a particular, non-singular minimal black hole model proposed by Hayward in Ref.~\cite{Hayward06}, which constructs a static spherically symmetric and asymptotically flat spacetime. {Recently, in Ref.~\cite{Babichev:2020qpr} this solution has been generalized to certain scalar-tensor theories, and new regular black holes have been reported in Refs.~\cite{Cisterna:2020rkc,Cano:2020ezi,Cano:2020qhy}, in the context of quasi-topological electromagnetic theories.
In fact, the non-singular nature of this black hole has made it an interesting topic studying its thermodynamics. Accordingly, and in Ref.~\cite{fmv20}, the laws of black hole thermodynamics have been investigated for the Hayward black hole (HBH) through studying the relations between its dynamical parameters $\{r_s, l\}$ that define the state of the system. 
In this paper we construct the correct foliation of the thermodynamics manifold by using the Carath\'{e}odory's approach, for which, the appropriate Pfaffian form $\delta Q_{\mathrm{rev}}$, representing the infinitesimal heat exchanged reversibly, is taken into account. {We should, however, state that the present study aims at establishing a new method of analyzing black hole thermodynamics and still is not constrained by the experimental data.}
Although the method does not require a priori knowledge of any of the so-called laws of thermodynamics, we will use the already known results for physical quantities, such as entropy, temperature, ect.
Therefore, the adiabatic surfaces are obtained by solving the Cauchy problem associated to the Pfaffian equation $\delta Q_{\mathrm{rev}}=0$. 
}

{\section{The Hayward black hole spacetime}\label{sec:HBH}}

{The Hayward black hole spacetime, is given by the regular, non-singular, static spherically symmetric  metric
\begin{equation}\label{eq1}
	{\rm d}s^2=-f(r)\,{\rm d}(c t)^2+\frac{{\rm d}r^2}{f(r)}+r^2{\rm d}\theta^2+r^2 \sin^2\theta \,{\rm d}\phi^2, 
\end{equation}
in which, the lapse function $f(r)$ is given by \cite{Hayward06}
\begin{equation}\label{eq2} 
f(r)=1-\frac{r_s\, r^2}{r^3+r_s\, l^2},
\end{equation}
where $r_s = 2 G M/c^2$ is the radius of the Schwarzschild black hole (SBH) of mass $M$, and $l$ is the Hayward's parameter ($0\leq l <\infty$), so that for $l=0$, the SBH is regenerated.} The spacetime admits an event horizon, which is obtained by solving the cubic equation $f(r) = 0$, and is given by \cite{fmv20} 
\begin{equation}
{r_+}=r_s\left(\frac{1+2\cos \alpha}{3}\right) \equiv  r_s R_+, \label{eq17}
\end{equation}
where $\alpha(r_s, l)=1/3\, \arccos\left(1-2 l^2/l_e^2\right)$,
$l_e = 2 r_s/\sqrt{27}$, and $0\leq l < l_e$. 
In the case that $l=l_e$, the roots reduce to the two degenerate positive values, and the EHBH is obtained representing the {{thermodynamic limit}} of the black hole. Hence, $l_e$ is the extremal limit of the Hayward's parameter.

\section{The $\{r_s, l\}$ thermodynamics in Carath\'{e}odory's approach }\label{termasp}

The most usual way to describe the thermo-geometric processes in black hole spacetimes, goes through the second law of black hole thermodynamics, which is postulated as
\cite{Bardeen:1973gs,Bekenstein:1972,Bekenstein:1973,Bekenstein:1974,Bekenstein:1975,Hawking:1974sw}
:
{The area of the black hole event horizon cannot decrease; it increases during most of the physical processes of the black hole,}
which, by means of the well-known B-H area-entropy formula
\begin{equation}
\label{entropy} 
S=\frac{k_B}{4}\frac{4 \pi r_+^2}{\ell_p^2},
\end{equation}
relates the entropy with the event horizon, where $k_B$ and $\ell_p$ are, respectively, the Boltzmann constant and the Planck length. Since the event horizon depends on the pair $\{r_s, l\}$, it is useful to define the {{metric entropy}} function
\begin{equation}\label{entrfcn}
\mathcal{S}(r_s, l)\equiv \frac{\ell_p^2\,S}{\pi k_B}=r_s^2\,R_+^2(r_s, l).
\end{equation}
On the other hand, the thermodynamics of the system can be approached, geometrically, in the context of the Carath\'{e}odory's approach, which
postulates the integrability of the Pfaffian form 
\begin{equation}\label{pf1}
   \delta Q_{{\rm rev}} = \mathcal{T} {\rm d} \mathcal{S},
\end{equation}representing the infinitesimal heat exchanged reversibly \cite{chandrasekhar39,Belgiorno:2002iw,Belgiorno:2002pm,Belgiorno_2003a,Belgiorno_2003b},
and this way, as highlighted in Ref. \cite{Belgiorno:2002iv}, it is connected with the Gibbs's thermodynamics.
Here $\mathcal{T}$ is the integrating factor representing the { absolute temperature}, defined by \begin{equation}\label{temp}
    \frac{1}{\mathcal{T}}=\left(\frac{\partial \mathcal{S}}{\partial r_s}\right)_l>0,
\end{equation}
which, by applying Eq.~\eqref{entrfcn}, yields \cite{fmv20}
\begin{equation}\label{teml2}
   \mathcal{T}=\frac{\mathcal{T}_s}{ R_+(\alpha)\left[R_+(\alpha)+ g_*(\alpha)\right]},
\end{equation}
where $\mathcal{T}_s\equiv (2r_s)^{-1}\equiv (4\mathcal{S}_s)^{-\frac{1}{2}}$ is the SBH temperature, and $g_*(\alpha) = \frac{4}{9}\sin\alpha \left(\csc 3\alpha-\cot 3\alpha \right)$.

Therefore, considering $\{r_s, l\}$ as independent thermodynamic coordinates,  the homogeneity of the system is reflected by the integrable Pfaffian form
\begin{equation}\label{pf01}
    \delta Q_{{\rm rev}} = {\rm d} r_s-\mathcal{F}_H {\rm d} l. 
\end{equation}
Note that, d$r_s$ and $-\mathcal{F}_H {\rm d} l$ represent, respectively, the internal energy and work, and the intensive variable (homogeneous of degree zero)
\begin{equation}\label{hf}
    \mathcal{F}_H=\frac{g_*(\alpha)}{R_+(\alpha)+g_*(\alpha)} \frac{r_s}{l},
\end{equation}
was introduced as the {{generalized Hayward's force}} in Ref.~\cite{fmv20}.
Here, an equilibrium geometrical state is compared with the equilibrium states of standard thermodynamics, by taking the infinitesimal variation of $\mathcal{F}_H$ in Eq. (\ref{pf01}), along the stationary HBH solution.
Then, the open non-extremal manifold $l < l_e$ corresponds to the thermodynamic domain, and is encompassed by the extremal sub-manifold (thermodynamic limit $\mathcal{T}=0$), formed by $l=l_e$. The foliation of this thermodynamic manifold can be generated by the integrability property 
$\delta Q_{{\rm rev}} \wedge {\rm d}(\delta Q_{{\rm rev}})=0$ (which is trivial in a two-variable case), specifically, on the submanifolds of codimension one, which are solutions of the Pfaffian equation 
$\delta Q_{{\rm rev}}=0$ (see the next sections).

\begin{figure}[h!]
    \begin{center}
    \includegraphics[width=7cm]{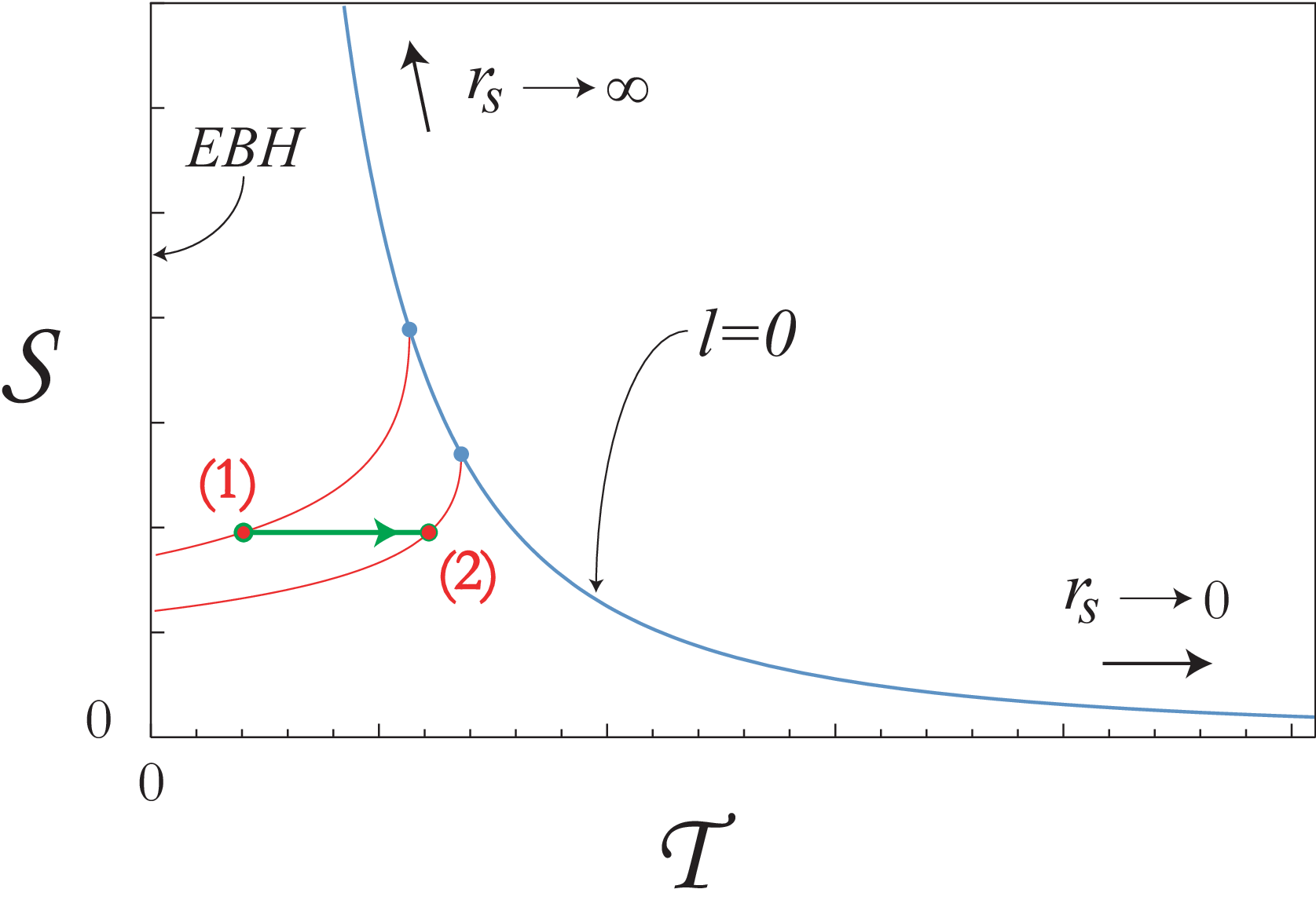}
    \end{center}
    \caption{{The $\mathcal{T}$-$\mathcal{S}$ diagram of the HBH, indicating both the EHBH and the SBH limits.
    }}
   \label{fts}
\end{figure}

Before continuing, let us turn our attention to Fig.~\ref{fts}, which shows the behaviors of the metric entropy and temperature. 
The physically accepted segment, lies within the domain $0\leq\mathcal{T}\leq \mathcal{T}_s$ (the blue curve). Note that, if $r_s$ is fixed, then $\Delta \mathcal{S} >0$ implies $\Delta l <0$. Therefore, by varying $l$ while keeping $r_s$ fixed (the red curves), the HBH transits to the SBH. It is also straightforward to see that, going from state (1) to state (2) (for which, $\mathcal{T}_1<\mathcal{T}_2$), the variable $r_s$ decreases. We can  therefore infer that, in an adiabatic process (the green arrow), both of the variables $(r_s,l)$ decrease simultaneously.

\section{The adiabatic processes and the extremal limit}\label{sec:extremal-thermo}

Letting $\mathfrak{r}_s\equiv r^{e}_s = \mathcal{F}_H^{e} \,l$
to be the extremal limit of $r_s$, where $\mathcal{F}_H^{e}=\sqrt{27}/2$ is the Hayward's force for the EHBH, one gets
\begin{equation}\label{extrconst2}
    \ed\mathfrak{r}_s=\mathcal{F}_H^{e}\,{\ed}l.
\end{equation}
Then, the area for the extremal states become
\begin{equation}\label{aext}
    \mathcal{A}_{e}=4 \pi \left(r^{e}_+\right)^2=4 \pi \left(r^{e}_s R^{e}_+\right)^2=\frac{16}{9} \pi  \mathfrak{r}_s^2,
\end{equation} which implies that 
\begin{equation}\label{daext}
    \ed\mathcal{A}_e = 24\pi l \ed l.  
\end{equation}
Thus, the {isoareal} condition $\ed\mathcal{A}_{e}=0$ is satisfied only if $l = \mathrm{const.}$, but these states still satisfy the Pfaffian equation
$\delta Q_{{\rev}}=\ed\mathfrak{r}_s-\mathcal{F}_H^{e}\,{\ed}l=0$.
Consequently, the adiabatic transformations are not isoareal transformations on the extremal submanifold.
We will return to this point later, but for now, the EHBH is regarded as an extremal submanifold that resides in the (adiabatic) integral manifold of $\delta Q_{{\rm rev}}$ in Eq.~\eqref{pf01}.

For non-extremal states the situation is different, since it is possible to obtain solutions for the isoareal equation $\ed\mathcal{A}=0$, and therefore, one can generate a foliation of the parameter space of the HBH, whose leaves are the surfaces $\mathcal{A}=\mathrm{const}.$ 
In fact, the Carath\'{e}odory's approach allows for foliating the thermodynamic manifold by means of the solutions to the Pfaffian equation $\delta Q_{{\rm rev}} = 0$, that provide a smooth and continuous 1-form field residing in the non-extremal sub-manifold. Accordingly, the integral manifolds of $\delta Q_{{\rm rev}}$ are surfaces with constant $\mathcal{S}$, which together with the paths that solve the Pfaffian equation, construct an {{isentropic}} surface (i.e. adiabatic and reversible) \cite{Belgiorno:2002pm}.  To elaborate on this point, let us apply the changes of variables  $x \doteq r_s^2$ and $y\doteq (\mathcal{F}_H^{e})^2 l^2$, so that Eq.~\eqref{pf01} can be recast as 
\begin{equation}\label{pf2}
    \delta Q_{{\rm rev}}=\frac{\ed x}{2\sqrt{x}} -\frac{\mathcal{F}_H(x, y)}{\mathcal{F}_H^{e}}\frac{\ed y}{2\sqrt{y}},
\end{equation}
which holds as long as $y\leq x$. 
Accordingly, the states which are connected adiabatically with the initial black hole state $(x_0, y_0)$, are solutions to the Cauchy problem
\begin{subequations}
\begin{align}
    &  \frac{{\rm d}y}{{\rm d}x} = \sqrt{\frac{y}{x}} \frac{\mathcal{F}_H^{e}}{\mathcal{F}_H(x, y)}, \label{pf3}\\
    &  y(x_0) = y_0, \label{pf4} 
\end{align}
\end{subequations}
where $x_0 > y_0$. Applying Eq.~\eqref{hf}, one can rewrite Eq.~\eqref{pf3} as
\begin{eqnarray}\nonumber
    \frac{\ed y}{\ed x} &=& \frac{y}{x}\left(1+\frac{R_+(x, y)}{g_*(x, y)}\right),\\
    \label{eq:cauchy}
   &=&\frac{1}{8}\left\{1+2\cos\left[\frac{1}{3}\arccos\left(1-\frac{2 y}{x}\right) \right] \right\}^3.
\end{eqnarray}
The above problem allows for two solutions, say $y_{1,2}$, which are given by 
\begin{equation} \label{is5}
   y_1(x) = x - y_2(x), 
\end{equation}
in which
\begin{equation}\label{is6}
   y_2(x) = \frac{27\rho^2 }{16}\left(1-\frac{\rho}{2 \sqrt{x}}\right),
\end{equation}
where $\rho$ is a constant determined by the initial condition. For extremal initial states $y(x_0)=x_0$, the Cauchy problem admits the two solutions $\rho=0$, and $\rho=2\sqrt{x_0}$. The first one corresponds to the adiabatic transformations between the extremal states, $y(x)=x$. As we have shown, these types of transformations are not isoareal. 
This statement indicates that the postulate $\mathcal{S} \propto \mathcal{A}$ is not valid for extremal black holes, and in particular, the EHBH.
The second solution is more complicated because it connects, adiabatically, the extremal states with non-extremal states.
Geometrically, this implies change in the topology and, due to the reversibility, problems with the second and the third laws of thermodynamics. 
To prevent any inconsistencies, it is better to eliminate this kind of solution.

\begin{figure}[t]
    \centering
    \includegraphics[width=6.9cm]{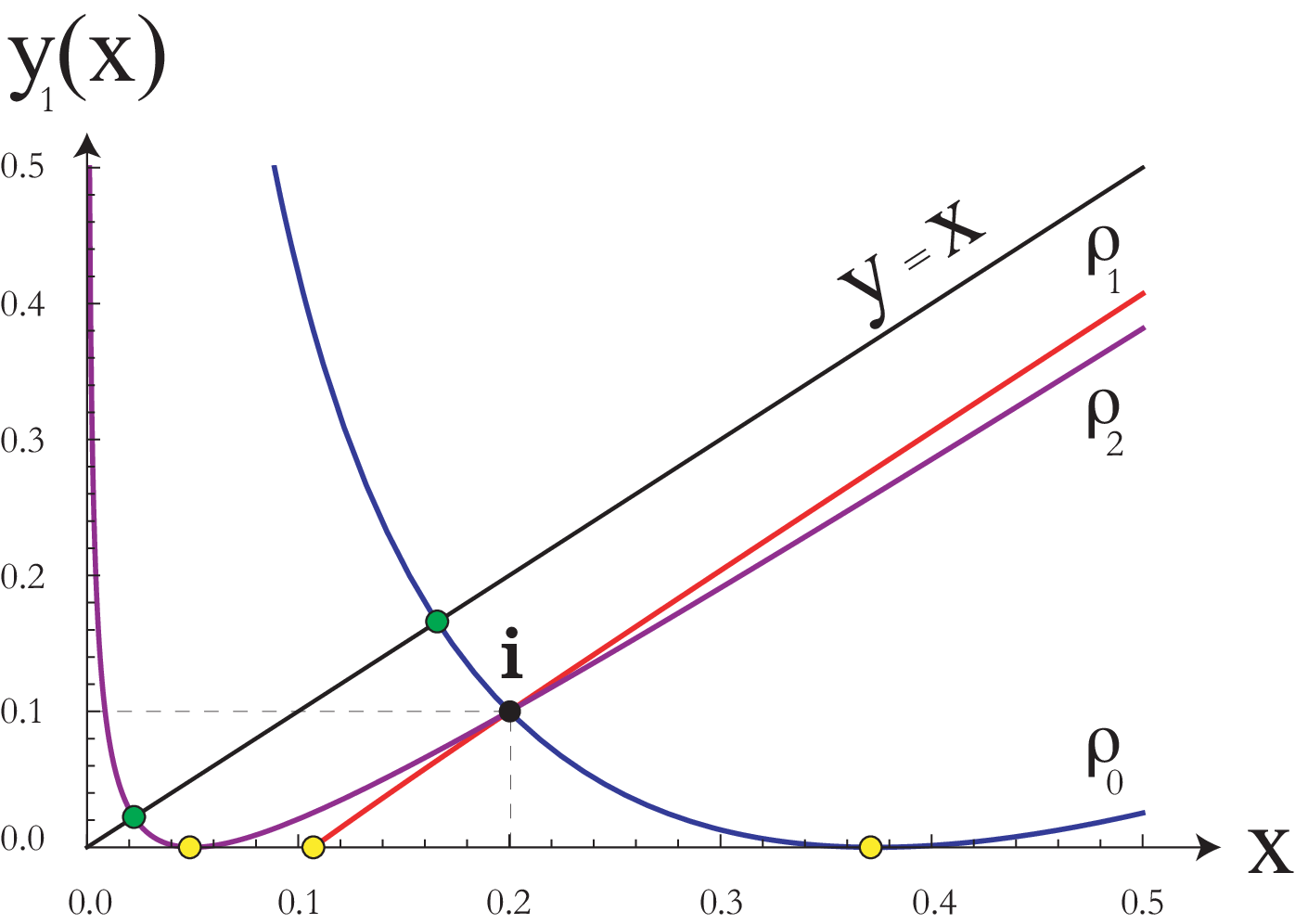}
    \includegraphics[width=6.9cm]{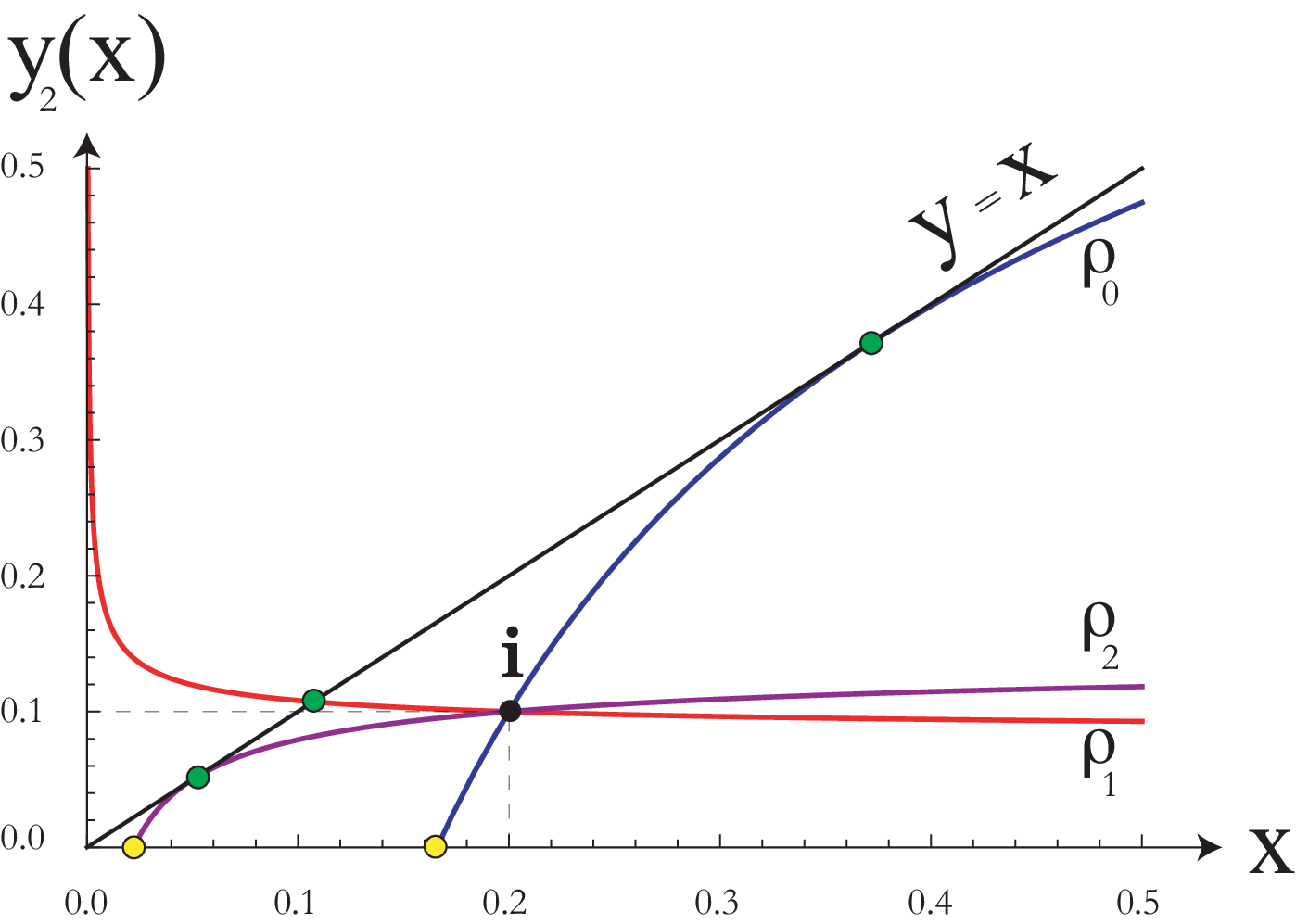}
    \caption{Adiabatic solutions for the Cauchy problem given by Eq. (\ref{eq:cauchy}). Top panel: $y_1(x)$ given by Eq. (\ref{is5}); Bottom panel:  $y_2(x)$ given by Eq. (\ref{is6}).
    In both plots we have used as the initial state $\mathbf{i}=(x_0, y_0)= (0.2, 0.1)$, so that $\rho_0= 0.815$, $\rho_1= -0.218$ and $\rho_2= 0.298$. Green dots represent the intersection of each function with the EHBH, $y(x)=x$, whereas the yellow represent the intersection with the SBH, $y(x)=0$.
    }
    \label{fy1}
\end{figure}

For $\rho \neq \{0, 2\sqrt{x_0}\}$ and considering an arbitrary, non-extremal, initial state, the functions given by Eqs. (\ref{is5}) and (\ref{is6}), yield the following equations for $\rho$:
\begin{eqnarray}\label{rhocub}
&&\rho^3-2\sqrt{x_0}\,\rho^2+\frac{32}{27}\sqrt{x_0}(x_0-y_0) = 0,\label{rhocub1}
\\
&& \rho^3-2\sqrt{x_0} \,\rho^2+\frac{32}{27}\sqrt{x_0}\, y_0 = 0,
\end{eqnarray}
whose solutions can be written simply as
\begin{equation}\label{is2}
\rho_k(x_0, y_0)=\frac{2\sqrt{x_0}}{3}\left[1+2\cos \left(\omega+\frac{2 k \pi}{3}\right)\right],
\end{equation}
where $k=0, 1, 2$, and 
\begin{equation}
    \label{is3}\omega\equiv \omega(x_0, y_0)=\frac{1}{3}\arccos\left|1-\frac{2 y_0}{x_0}\right|.
\end{equation}
The above means that, given an initial equilibrium configuration for the HBH, there are six possible curves adiabatically connected with it, say $y_{j k}(x)\equiv y_j(x; \rho_k)$, with $j=1, 2$ and $k=0, 1, 2$. 
Let us designate by $\epsilon_{jk}$ and $\sigma_{jk}$, the value of the $x$-coordinate at the intersection of each curve $y_{jk}$ with the EHBH (where $y=x$), and the SBH (where $y=0$), respectively (green and yellow dots in Fig. \ref{fy1}). Then, it is no hard to show that $\epsilon_{1k}=\sigma_{2k}$ and $\epsilon_{2k}=\sigma_{1k}$. 
In addition, since $\rho_0>\rho_2>0$, one gets $\epsilon_{10}=\sigma_{20}>\epsilon_{12}=\sigma_{22}>0$, and consequently, $\epsilon_{20}=\sigma_{10}>\epsilon_{22}=\sigma_{12}>0$. However, because $\rho_1<0$, the function $y_2(x)$ is strictly positive and therefore $\epsilon_{21}=\sigma_{11}>0$ and $\epsilon_{11}=\sigma_{21}\rightarrow \infty$.

The above result is crucial to exclude the leaf $\mathcal{T}=0$ from the adiabatic manifold. In fact, if we delete all the solutions $y_{jk}$, except $y_{11}$, then the generated adiabatic surface does not intersect the extremal surface (see Fig. \ref{fy2}).
Thus, assuming that the relation $\mathcal{S} \propto \mathcal{A}$ is not followed by the extremal black holes \cite{Teitelboim:1994}, we characterize the thermodynamics of the HBH by
\begin{equation}\label{eq:A40}
  \mathcal{S}(r_s, l) = \left\{
             \begin{array}{ll}
             \mathcal{A}/4,&\, \textrm{non-extremal states,} \\
             \medskip\\
             0,&\, \textrm{extremal states.} \\
             \end{array} \right.
\end{equation}
To ensure the correct foliation of the thermodynamic manifold, and based on the processes described in the last part of Sect. \ref{sec:HBH} (cf. Fig. \ref{fts}), the following conditions must hold:
\begin{figure}[t]
    \centering
    \includegraphics[width=7.6cm]{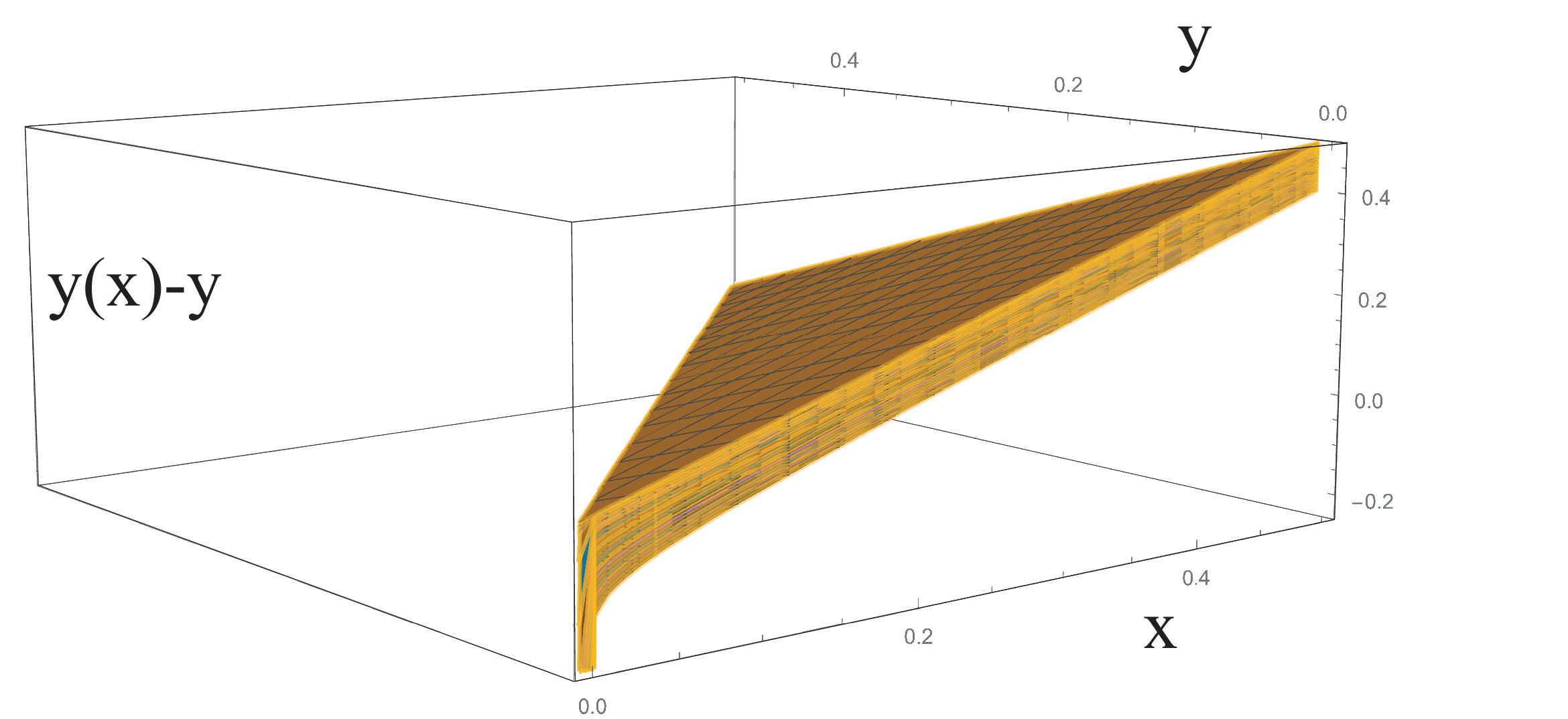}
    \caption{Adiabatic volume generated by the physical solution  $y(x)=y_{11}(x)$. The construction corresponds to the foliation of the surfaces $y(x)-y$  by fixing $x_0=0.2$ and varying $y_0$ between $0$ and $0.198$ (with the steps $0.002$). The upper limit edge of the surface corresponds to the extremal surface $x$-$y$.
    }
    \label{fy2}
\end{figure}
\begin{enumerate}
    \item The slope of the $x$-$y$ curves is positive,
    \begin{equation}\label{eq:cond1}
        \frac{\ed y}{\ed x} >0.
    \end{equation}
    \item\label{c2} As the variables decrease, the system evolves towards the SBH, while when they grow, the EHBH is approached.
    \item\label{c3}  In the neighborhood of any arbitrary state, $\mathbf{i}$, of a physical system there are neighboring states $\mathbf{i'}$, which are inaccessible from $\mathbf{i}$ along adiabatic paths ({ Carath\'{e}odory's principle} \cite{Caratheodory09,buchdahl49II}).
\end{enumerate}
Therefore, the complete solution to the Cauchy problem that satisfies the physical requirements, can be written as
\begin{equation}
    \label{slf}y(x)=x-\frac{27 \rho_1^2}{16}\left(1+\frac{|\rho_1|}{2\sqrt{x}}\right),
\end{equation}
whose asymptotic behavior is
\begin{equation}\label{asym}
    y(x) \simeq x-\frac{27 \rho_1^2}{16}, 
\end{equation}
that ensures the condition $x>y$.
Recently, an study on the adiabatic properties of the Bañados-Teitelboim-Zanelli (BTZ) black hole was carried in Ref. \cite{FLV21}, showing that the correct thermodynamic foliation is possible by building a piecewise argument, with a physically accepted segment, which is the solution (\ref{slf}).

\section{Classical scattering of two EHBHs and the second law}\label{sec:scatter}

Let us assume that the thermodynamic state $(x_a+x_b, y_a+y_b)$, is produced by merging two HBHs of the initial conditions $(x_a, y_a)$ and $(x_b, y_b)$, in a process that no exchange of energy is done with the rest of the universe. Defining the quantity
\begin{equation}\label{cs1}
	\zeta^{2} (x, y)=x-y, 
\end{equation}
the initial state of the process is now characterized by $\zeta_{\mathrm{in}}=\zeta_{a}+\zeta_{b}$, with $\zeta_{a}^{2}\equiv \zeta^{2}(x_a, y_a)\geq 0$ and $\zeta_{b}^{2}\equiv \zeta^{2}(x_b, y_b)\geq 0$. In the same manner, the final state becomes $\zeta_{\mathrm{fin}}=\zeta_{ab}$, where $\zeta _{ab}^{2}\equiv \zeta^{2}(x_a+x_b, y_a+y_b)$, and Eq.~\eqref{cs1} yields
\begin{eqnarray}\label{cs3}
\zeta_{ab}^2  &=&	\zeta_{a}^{2} + \zeta_{b}^{2} +2\left( x_a x_b - y_a y_b\right). 
\end{eqnarray}
Exploiting Eq.~\eqref{cs3}, we obtain 
\begin{equation}\label{fisc}
 \zeta_{ab}^{2} - (\zeta_a+ \zeta_{b})^{2}=2\Big[x_a x_b-y_a y_b -\sqrt{(x_a^2-y_a^2)(x_b^2-y_b^2)}\,\Big].
\end{equation}
In the case that the initial states are constituted by extremal black holes (with zero entropy according to Eq.~\eqref{eq:A40}), we have $x_a=y_a$ and $x_b = y_b$, giving $\zeta_a=0=\zeta_b$ and hence, $\zeta_{ab}=0$, which implies that the final black hole is as well, extremal and therefore, has zero entropy.

If the merger of the black holes is considered irreversible, then the process would violate the second law of thermodynamics, because they never produce a regular HBH to increase the total entropy of the system. Although this argument holds for systems at non-zero temperature, it puts into question the hypothesis $\mathcal{S} = 0$ for the extremal states. 
To relax the above, we could consider a small amount of angular momentum so that the final state is a non-extremal state, and thus, the final entropy is greater than the initial one, protecting the second law.
This obviously involves a detailed study of such a situation that will not be addressed here.
Another way that could alleviate the understanding of the process, consists of abandoning the hypothesis of zero entropy to give rise to a law of the form $\mathcal{S} = f(\mathcal{A})$, where $f$ is a function of the area, as proposed, for example, in Ref. \cite{Lemos:2017aol} in the context of a thin shell for the case of a rotating uncharged BTZ black hole.


{\section{Discussion and final remarks}\label{Summary}}

The correct construction of a thermodynamic manifold allows us establishing its known laws with complete property and thus, connecting this realization with the so-called thermo-mechanics of black holes.
This makes it possible to identify the isoareal and adiabatic transformations, since for non-extremal black holes, the law $\mathcal{S}\propto \mathcal{A}$ is valid.


In this letter we have constructed the correct foliation of the thermodynamic manifold for the HBH, by applying the Carath\'eodory axiomatic principle. Accordingly, we solved the Cauchy problem associated with the Pfaffian equation $ \delta Q_{\mathrm{rev}} = 0$ for the HBH, where $ \delta Q_{\rev}$ represents the infinitesimal heat exchanged reversibly. It is important to note that, even the procedure of applying the B-H formula, does not demand any {priori} knowledge of the laws of thermodynamics, in order to construct the submanifold, although of course, they are mathematically connected.

Developing the aforementioned ideas, we found that, given an initial state of equilibrium for the HBH, one can find twelve isoareal connections with other equilibrium states, through six possible curves, $y_{jk}$ with $j=1,2$ and $k=0, 1, 2$.
Five of these curves adiabatically connect the non-extremal states with the extremal ones, which causes contradictions with the second and the third laws; for example, it would be possible to get to the zero temperature in a finite sequence of steps, and therefore, build a heat engine whose efficiency is equal to one.
We, however, were able to find a solution that avoids such unwanted behavior; it generates a manifold that does not include the extremal sub-manifold $\mathcal {T} = 0$.
This can be better understood in the context of the Carath\'{e}odory's principle (cf. condition \ref{c3}), by assuming such states as being inaccessible.

On the extremal sub-manifold, the condition $\delta Q_{\rev} = 0 $ is still valid, which gives rise to two kinds of transformations that satisfy the initial extremal condition $y(x_0)=x_0$. One of the solutions, adiabatically connects the extremal states with the non-extremal ones, that have the same areas (isoareal transformations). The other solution connects the extremal states with different areas. This solution, however, can be considered as an adiabatic transformation, by virtue of the null entropy law.
This accounts for the disconnection between the leaves $\mathcal {T} = 0$ and $\mathcal {T} > 0$, which is expressed, as well, in the metric entropy (\ref{eq:A40}). In fact, as stated above, such disconnection can be provoked by the exclusion of the solutions $ y_ {jk} $ (except for $y_ {11}$), that allows for eliminating any connections between the above varieties, and respects the second and the third laws.

Nevertheless, the scattering process of two static EHBH leads to questioning this issue, even more profoundly, because of the impossibility of the increase in entropy in the final state. 
As in the other cases, the addition of a new variable seems to be the solution, although the price will be the loss of the homogeneity of the system, which gives rise to quasi-homogeneous potentials that do not allow for fixing the degree of homogeneity of the Pfaffian form, although, a thermodynamic construction is allowed \cite{Belgiorno:2002iw}.

{Finally, all the standard thermodynamic parameters can be connected with each other, in order to obtain different quantities which are of interest. For example, using the Stefan-Boltzmann law together with the corresponding Hayward's quantities, one can calculate the evaporation time of the HBH, from the differential equation
\begin{equation}\label{eq:evaporationdiff}
    \frac{\ed r_s}{\ed t}=-b r_s^2 R_+^2(r_s,l) \mathcal{T}^4(r_s,l),
\end{equation}
where $b=l_p^2 c/(120\pi)$. Expansion of the right hand side of Eq.~\eqref{eq:evaporationdiff} about the Schwarzschild solution (corresponding to $l=0$) and doing the intergation, yields
\begin{equation}\label{eq:evaporationint}
    \Delta t_\mathrm{evap} \simeq \frac{16 r_s^3}{3 b}\left(1+6 \frac{l^2}{r_s^2}\right),
\end{equation}
which implies that the lifetime of the HBH is longer than that of the SBH.
}

\section*{Acknowledgements}
M.F. has been supported by the Agencia Nacional de Investigaci\'{o}n y Desarrollo (ANID) through DOCTORADO Grants No. 2019-21190382, and No. 2021-242210002. J.R.V. is partially supported by Centro de Astrof\'isica de Valpara\'iso.


\appendix


\bibliographystyle{elsarticle-num-names}
\bibliography{biblio_v1.bib}


\end{document}